# GCVAMD: A Modified CausalVAE Model for Causal Age-related Macular Degeneration Risk Factor Detection and Prediction

Daeyoung, Kim

***Abstract*— Age-Related Macular Degeneration(AMD) has been one of the most leading causes of permanent vision impairment in ophthalmology. Though treatments, such as anti-VEGF drugs or photodynamic therapies, were developed to slow down the degenerative process of AMD, there is still no specific cure to reverse vision loss caused by AMD. Thus, for AMD, detecting existence of risk factors of AMD or AMD itself within the patient's retina in early stages is a crucial task to reduce the possibility of vision impairment. Apart from traditional approaches, deep learning based methods, especially attention-mechanism based CNNs and GradCAM based XAI analysis on OCT scans, exhibited successful performance in distinguishing AMD retina from normal retinas, making it possible to use AI-driven models to aid medical diagnosis and analysis by ophthalmologists regarding AMD. However, though having significant success, previous works mostly focused on prediction performance itself, not pathologies or underlying causal mechanisms of AMD, which can prohibit intervention analysis on specific factors or even lead to less reliable decisions. Thus, this paper introduces a novel causal AMD analysis model: GCVAMD, which incorporates a modified CausalVAE approach that can extract latent causal factors from only raw OCT images. By considering causality in AMD detection, GCVAMD enables causal inference such as treatment simulation or intervention analysis regarding major risk factors: drusen and neovascularization, while returning informative latent causal features that can enhance downstream tasks. Results show that through GCVAMD, drusen status and neovascularization status can be identified with AMD causal mechanisms in GCVAMD latent spaces, which can in turn be used for various tasks from AMD detection(classification) to intervention analysis.**

## I. Introduction

Age Related Macular Degeneration, or AMD, is considered as one of the most leading factors of human vision impairment. According to the CDC and NHANES-based estimates, it was found that approximately 18.34 million were under early stage AMD in the US (2019), while 1.49 million were found to suffer from late-stage AMD (US, 2019)[1][2]. Aside from being a leading cause, AMD itself has high significance due to its risk of causing permanent vision loss [3]. Though there exists some treatments such as anti-VEGF drugs or photodynamic therapies for wet AMD [4], most treatments are only successful in slowing down the speed of vision impairment, not in reversing the loss of vision. Moreover, it is found that there is still no cure for late dry AMD, which generally leads to significant damage in the central vision and lighting in patients. Thus, in current aspects, it is crucial to find a way to detect AMD in early stages of degeneration with precision to maximize the possibility of deceleration of vision loss and enhance the welfare of patients. In traditional AMD detections, visual interpretations by human ophthalmologists were mainly driven based on OCT images or fluorescein angiography images of patients to track optical component abnormalities. While this approach is still considered valid, under massive advancements in computer vision models, artificial intelligence-based diagnosis of AMD with retinal image patient data is also currently widely being implemented in various fields to aid or enhance possibility of accurate AMD detection. For example, Convolutional Neural Network(CNN) with attention mechanisms and image-based XAI methods exhibit significant success in AMD image classification performance. Specifically, implementing multi-level dual attention mechanisms in CNN architectures, such as ResNet, robust high performing AI-driven AMD detection models were achieved with OCT images of patients in [5,6], when compared to previous methods such as HOG-SVM or VGG. Meanwhile, in researches such as [7,8], fine-tuning GoogLeNet with transfer learning under ImageNet data, or connecting ophthalmologist-based domain knowledge with ResNet50 encoding + K-means clustering for AMD risk biomarker detections exhibited significant success in enhancing AMD classification accuracy and risk evaluation. On the other hand, researches such as [9,10] achieved XAI-based interpretability in AMD retinal image analysis through combining GradCAM with fine-trained VGG16 models, or combining guided backpropagation with OCT/Infrared Reflectance(IR) based CNN models.

*Affiliation Information: The Department of Applied Statistics, Yonsei University, Seoul, South Korea.
*First Author, Corresponding Author. (Email: lgtlsafg@yonsei.ac.kr)

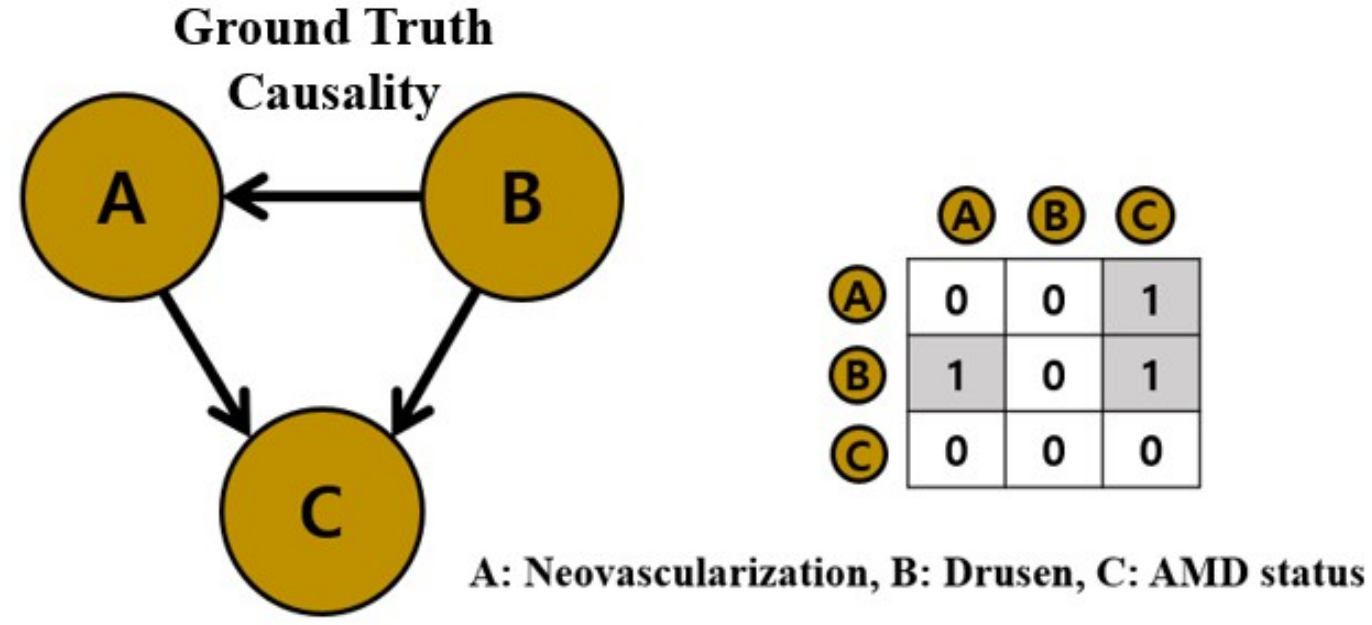

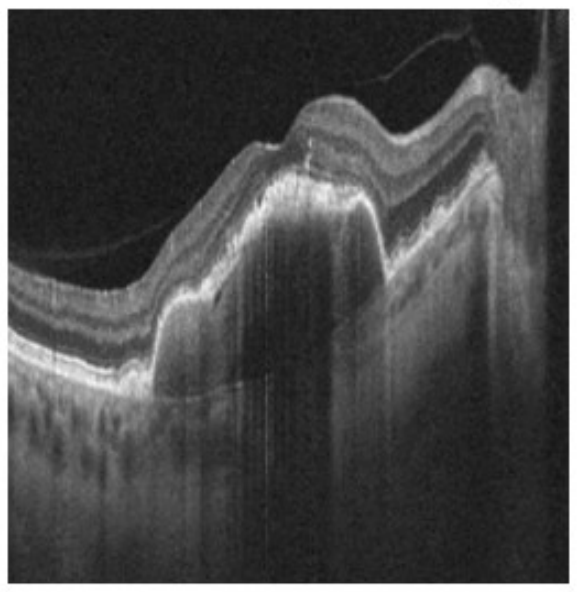

**Fig 1. (Left) Visualization of ground truth AMD causality with adjacency matrix, (Right) OCT image re-labelling example**

However, though significant success in detecting AMD under neural networks has been repetitively reported in various research, there still remains issues. As most of previous research focused on prediction performance itself, resulting deep learning models only considered correlations between variables, which can diverge from ground truth causal AMD structures. Thus, intervention analysis in terms of treatments may not be applicable in current AMD detection models. Moreover, reliability issues can also occur due to strong reliance on correlations. This research therefore attempts to achieve a causality-based AMD detection model which can enable (i) causal intervention based inference, (ii) detection of causal factors of AMD, and (iii) enough valid prediction performance that can alternate conventional CNN based approaches. Using OCT images from normal/AMD patients and causal representation learning mechanisms such as CausalVAE [11], results show that valid detection of causal factors of AMD, such as drusen and neovascularization, within OCT images and success in returning reliable latent causal structures for intervention analysis can be achieved through the proposed framework. Furthermore, with causal latent factors as new inputs, reliable, yet enhanced downstream DL models were found to be achieved regarding classification of AMD/Normal retinas.

## II. Preliminaries

### A. Data

For analysis, OCTDL, an OCT open access dataset created by M, Kulyabin et al.(2023), was implemented [12]. OCTDL dataset consists of 2064 B-scan images, which were acquired through Optovue Avanti RTVue XR with axial and transverse resolutions each being $5\mu$m and $15\mu$m. Each scan being labelled by clinical specialists under initial scanning by a group of 7 medical students, 1231 AMD OCT images, 332 Normal OCT images, 147 Diabetic Macular Edema images, 155 Epiretinal Membrane images, 22 Retinal Artery Occlusion images, 101 Retinal Vein Occlusion images and 76 Vitreomacular Interface Disease images were deduced for analysis. Furthermore, each image was linked with a meta data file which contains subcategories or conditions of disease, patient ID, or scales of images. Specifically, for AMD OCT images, causes of AMD such as Macular Neovascularization or Drusen, and intensities of AMD were included within corresponding meta data files.

In this work, to focus on extracting causal representations of AMD, 1231 OCT images of the AMD retina and 332 images of normal retina with corresponding metadata were implemented for model fittings. Furthermore, for causal identification, two additional variables, which each denote Drusen existence(0: No, 1: Yes) and Neovascularization existence (0: No, 1: Suspected, 1: Yes), were defined for experiments based on meta data. Using additional variables and an AMD status variable, which was mapped to 0 to 3 by its severity(0: None, 1: Early, 2: Intermediate, 3: Late), as causal indicator variables, AMD causal representation learning was computed and evaluated in this research. Specifically, under domain ophthalmologic knowledge from [13-15], ground truth AMD causal graph and its corresponding causal adjacency matrix was set as the comparison criterion for causal structure learning (Fig 1). In terms of dataset size, 150 random AMD image data and 150 random normal image data was implemented in causal learning to contemplate high label imbalance present in OCTDL datasets. Lastly, in terms of image size, every original OCT data was converted into 224 x 224 format for analysis.

### B. Causal Discovery

For decades, extracting causal information from observational data has been explored in various aspects. One main branch within the field of causal discovery has been the Structural Causal Model (SCM) based causality extraction approach, which was first dealt by J, Pearl [16,17]. Expanding the concept of Structural Equation Models(SEM) in Economics to the field of causal framework, J, Pearl assumed equation (1) can represent both the causal dependency structures and effects of causal interactions within data. Based on this assumption, conditional independence-based causal discovery algorithms, such as PC-algorithms or IAMB, and score-based algorithms such as Hill-climbing algorithms or Min-Max Hill climbing algorithms, had been implemented to detect causality between observational variables [18].

$$\boldsymbol{X} = A^T\boldsymbol{X} + \boldsymbol{\varepsilon} \ \ s.t.\ \ \mathcal{G} = \langle A, X \rangle, \boldsymbol{\varepsilon}\text{: noise variable, A: adj mat} \qquad \textbf{(1)}$$

However, due to high inefficiency in independence test approaches and the burden caused by non-continuous causal graph search in most score-based approaches, continuous causal graph learning, or continuous causal structure learning has been a major topic in recent fields of causal discovery, leading to the novel algorithms such as NOTEARS, DAG-GNN, or Graph Autoencoders. For example, NOTEARS algorithm assumes that the components in the adjacency matrix can be considered as learnable weight parameters in neural networks, making it possible to update adjacency components under gradient descent algorithms, with directed acyclic graph(DAG) constraints converted to augmented lagrangian components [19]. Meanwhile, DAG-GNN algorithms encode VAE inputs into noise variables in SCMs under the assumption of (2). Using this approach, DAG-GNN incorporates both non-linearity and continuous structure learning in causal DAG extractions under gradient descent algorithms, which exhibited superior performance in detecting various causal structures within tabular data [20]. On the other hand, Graph Autoencoder based causal discovery algorithms extract causal connections within data under the assumption of expanding NOTEARS to non-linear spaces as in equation (3), leading to flexible, yet effective SCM fittings in numerous causal data analysis [21].

$$\boldsymbol{X} = A^T\boldsymbol{X} + \boldsymbol{\varepsilon} \iff (\boldsymbol{I} - \boldsymbol{A^T})\boldsymbol{X} = \boldsymbol{\varepsilon}$$
$$\Rightarrow \boldsymbol{f_2}((\boldsymbol{I} - \boldsymbol{A^T})\boldsymbol{f_1}(\boldsymbol{X})) = \boldsymbol{\varepsilon} \ \ (non - linear) \tag{2}$$

$$\boldsymbol{X} = \ f_2\left(\boldsymbol{A^T} f_1(\boldsymbol{X})\right) + \boldsymbol{\varepsilon}, \ \ f_{1(2)} \approx Neural\ Network_{1(2)} \tag{3}$$

Based on the success of continuous causal structure learning in various research, current causal discovery research is mainly focused on extracting latent causal information from low-level observational data using continuous adjacency matrix learning, which is called "Causal Representation Learning" in most frameworks. Representative Causal Representation Learning (CRL) methods, such as CausalVAE [11], diffusion based CRL [22], or weakly supervised CRL in [23], all exhibit that under certain constraints, underlying causal features can be retrieved from image data with only neural network based approaches on hand. For example, CausalVAE expands the framework of DAG-GNN encoders to Convolutional VAE latent space extractions, while adding a structural causal layer that updates the adjacency matrix as in NOTEARS. This led to success in retrieving ground truth causal structures within *CelebA* [24], *Pendulum* and *Flow* causal datasets [11]. Based on current success in CRL, this research implemented a CausalVAE + Graph Autoencoder approach to extract and identify latent AMD related causal factors, while enabling neural networks to perform intervention analysis with valid causal structure information.

## III. METHODOLOGY

### A. GCVAMD framework for causal AMD detection

This research introduced a novel causal AMD detection model for advanced AI-driven medical inference and diagnosis: '**G**raph autoencoder based **C**ausal**VAE** for **AMD** detection' (GCVAMD). GCVAMD implements the CausalVAE approach to extract latent causal factors within the OCT retinal image, while assuming non-linear SCM assumptions based on Graph Autoencoder structures for realistic approaches. That is, while CausalVAE assumes a latent linear SCM for causal inference, GCVAMD assumes non-linearity for both latent factor Z retrieval from noise variables: $\varepsilon$ and causal links between latent variables, *Z*. For further explanation, basic information regarding frameworks of CausalVAE and Graph Autoencoders were summarized as follows.

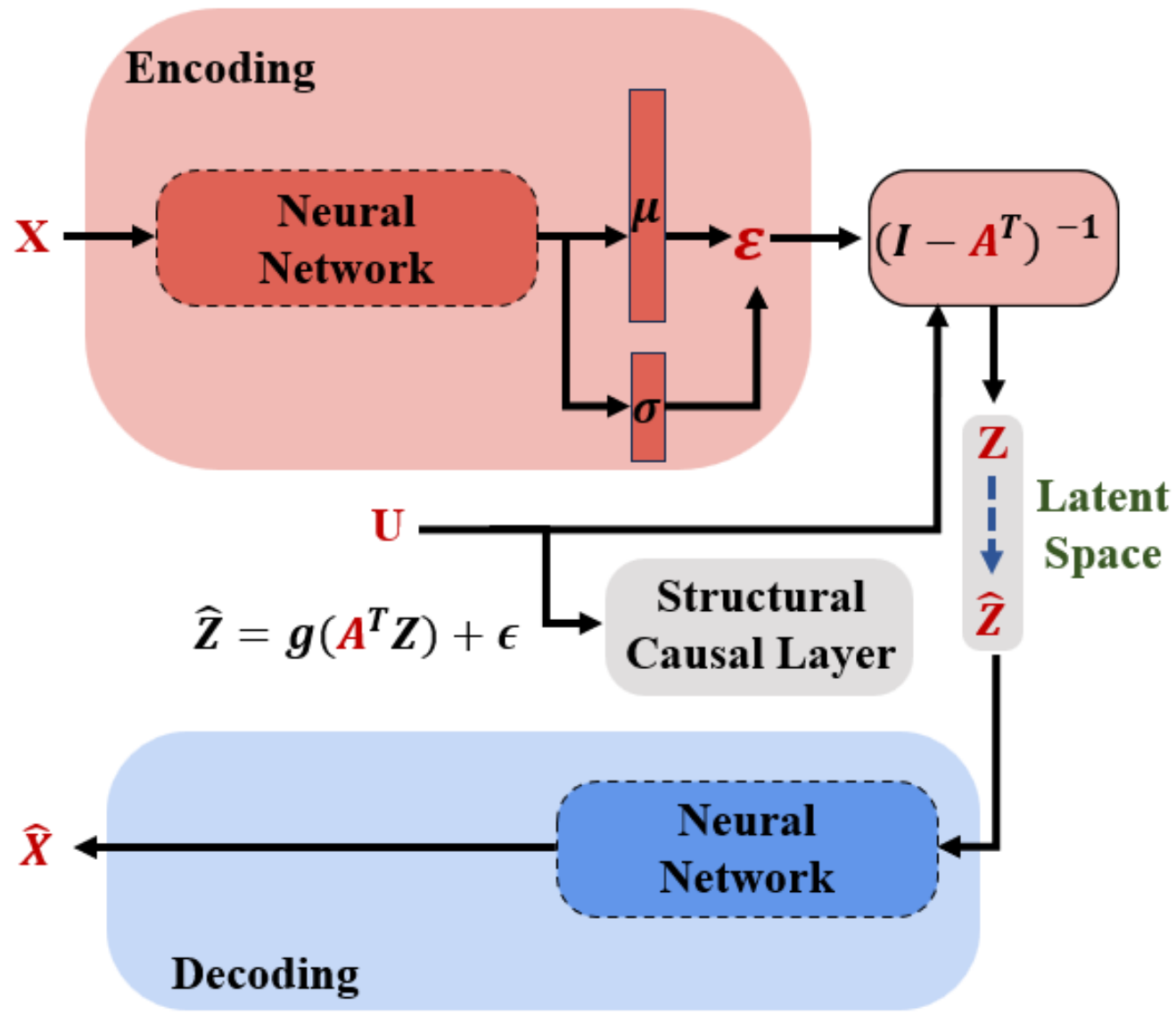

**Fig 2. Visualization of basic CausalVAE framework in [11].**

#### (1) CausalVAE

CausalVAE is a Variational Autoencoder based latent causal representation learning framework designed for image data analysis. Based on equation (4), CausalVAE first (i) encodes noise variables within the SCM, and (ii) compute latent factor Z under a learnable weighted adjacency matrix *A*. Next, latent vector *Z* in CausalVAE goes through a (iii) structural causal layer in (5), which can be considered a generalized additive model approach for SCM computation under masking (Fig 2).

$$(\boldsymbol{I} - \boldsymbol{A^T})\boldsymbol{Z} = \boldsymbol{\varepsilon} \ \Rightarrow (\boldsymbol{I} - \boldsymbol{A^T})^{-1}\boldsymbol{\varepsilon} = \ \boldsymbol{Z} \tag{4}$$
$$for\ \forall i^{th}\ variable\ in\ \boldsymbol{Z}, \ \ \ \boldsymbol{\hat{Z}_i} = \ g_i(\boldsymbol{A_i}, \boldsymbol{Z};\ \boldsymbol{\theta_i}) + \boldsymbol{\epsilon_i} \tag{5}$$

Lastly, (iv) reconstructed vector $\hat{Z}$ becomes the input for the convolutional Decoder, which, in turn, is used to reconstruct original input images. Using additional label information variables *U*, latent causal variables are supervised to correspond with domain labels which contain causal information. Thus, loss function of CausalVAE is comprised of VAE Loss, latent vector Z reconstruction loss and label variable U loss, with DAG conditions considered as constraints of optimization.

$$
\begin{aligned}
&-\log\big(p(x|u)\big) = -\int \log\big(p(x|u)\big)\, q_\phi(z,\varepsilon|x,u)dzd\varepsilon \\
&= -\int \log\Big(\frac{p(x,z,\varepsilon|u)}{p(z,\varepsilon|x,u)}\Big)\, q_\phi(z,\varepsilon|x,u)dzd\varepsilon \\
&= -\int \log\left(\frac{p(x,z,\varepsilon|u)}{q_\phi(z,\varepsilon|x,u)}\right) q_\phi(z,\varepsilon|x,u)dzd\varepsilon \\
&\quad -\int \log\left(\frac{q_\phi(z,\varepsilon|x,u)}{p(z,\varepsilon|x,u)}\right) q_\phi(z,\varepsilon|x,u)dzd\varepsilon \\
&= -ELBO(\phi) - KL(q_\phi(z,\varepsilon|x,u)\ ||\ p(z,\varepsilon|x,u)) \\
&\quad (\leq -ELBO(\phi), \because KL(constant) \geq 0) \qquad \textbf{(6)}
\end{aligned}
$$

$$
\begin{aligned}
&= -ELBO(\phi) = -\int \log\left(\frac{p(x,z,\varepsilon|u)}{q_\phi(z,\varepsilon|x,u)}\right) q_\phi(z,\varepsilon|x,u)dzd\varepsilon \\
&= -\int \log\left(\frac{p(x|u,z,\varepsilon)p(z,\varepsilon|u)}{q_\phi(z,\varepsilon|x,u)}\right) q_\phi(z,\varepsilon|x,u)dzd\varepsilon \\
&= -\int \log(p(x|u,z,\varepsilon))\, q_\phi(z,\varepsilon|x,u)dzd\varepsilon \\
&\quad +\int \log\left(\frac{q_\phi(z,\varepsilon|x,u)}{p(z,\varepsilon|u)}\right) q_\phi(z,\varepsilon|x,u)dzd\varepsilon \\
&= -E_q\big(\log\big(p(x|u,z,\varepsilon)\big)\big) + KL(q_\phi(z,\varepsilon|x,u)\ ||\ p(z,\varepsilon|u))
\end{aligned}
$$

$$
\begin{aligned}
&\therefore ELBO = E_q\big(\log\big(p(x|z)\big) - KL(q_\phi(z,\varepsilon|x,u)\ ||\ p(z,\varepsilon|u)\big) \\
&= E_q\big(\log\big(p(x|z)\big) - KL(q_\phi(\varepsilon|x,u)\ ||\ p(\varepsilon)) - \\
&KL(q_\phi(z|x,u)\ ||\ p(z|u)) \qquad \textbf{(7)}
\end{aligned}
$$

Specifically, VAE loss is summarized into Binary Cross Entropy loss and a KL-divergence loss based on proof in (6) to (8) (*L=1 assumed), whereas Z and U reconstruction losses are each defined as MSE loss. Thus, total CausalVAE loss is defined as (9). Here, DAG constraint H($A$) is an equivalent representation of (10).

$$
\begin{aligned}
-ELBO &\simeq -\frac{1}{L}\sum_l \log\left(p\left(\boldsymbol{x}|\boldsymbol{z^l}\right)\right) + \frac{1}{2}\sum {\varepsilon_{ij}}^2 \\
&\quad + KL(q_\phi(z|x,u)\ ||\ p(z|u)) \\
&= -\log\prod_{j=1}^{D} p\left(x^{(j)}|\boldsymbol{z^l}\right) + \frac{1}{2}\sum {\varepsilon_{ij}}^2 \\
&\quad + KL(q_\phi(z|x,u)\ ||\ p(z|u)), (L=1) \\
&= BCE\left(p_{ij}, x_{ij}\right) + \tfrac{1}{2}\sum {\varepsilon_{ij}}^2 + KL(q_\phi(z|x,u)\ ||\ p(z|u)) \qquad \textbf{(8)}
\end{aligned}
$$

$$
\begin{gathered}
CausalVAE\ loss \\
L = -ELBO + \alpha H(A) + \beta l_u + \gamma l_z \\
* l_u = \sum \|u - A^T u\|^2, l_z = \sum \|z - g(A^T z)\|^2 \\
*H(A) = tr\left(\left(I + \tfrac{c}{m} A \circ A\right)^d\right) - d \qquad \textbf{(9)}
\end{gathered}
$$

$$
H(A) = tr\left(e^{A \odot A}\right) - d, A \in \mathbb{R}^{d \times d} \qquad \textbf{(10)}
$$

In this research, equation (10) was used for DAG constraint incorporation. Furthermore, considering the fact that mis-labelling issues can exist within the OCTDL dataset metadata, and the fact that three label information: Neovascularization Status {0,1}, Drusen Status {0,1}, and Severity of AMD {0,1,2,3} defined in *II. Preliminaries* have diverse (non-identical) discrete label spaces, which can cause non-standardization related issues, KL-divergence of $KL(q_\phi(z|x,u)\ ||\ p(z|u))$ in (7) was replaced by a weighted mean squared error (MSE) between Z and U: $v \times MSE(Z,U)$, which can alleviate the possibility of excessive supervision in finding valid AMD causal representations. In further discussions, $MSE(Z,U)$ will be denoted as $l_{zu}$.

*(2) Graph Autoencoder*

Graph Autoencoder in causal frameworks denotes a continuous causal adjacency matrix learning method based on non-linearity assumptions as in equation (3). [21]

$$
\boldsymbol{X_i} = g_{2,i}\left(\boldsymbol{A_i}^{\boldsymbol{T}} g_{1,i}(\boldsymbol{X})\right) + \boldsymbol{\varepsilon},\ \ g_{1(2)} \approx Neural\ Network_{1(2)}
$$

Unlike the Structural Causal Masking layer introduced in CausalVAE, which only implements a non-linear function once for $A_i$ and factors Z, Graph Autoencoder based approaches implement two separate MLPs to compute highly complex causal structures within variables (Fig 3). By setting functions: $g_1$ and $g_2$ individually for each $X_i$ under masking, generalized additive MLPs can represent the causal flow in accordance with the original assumption of J, Pearl's SCM.

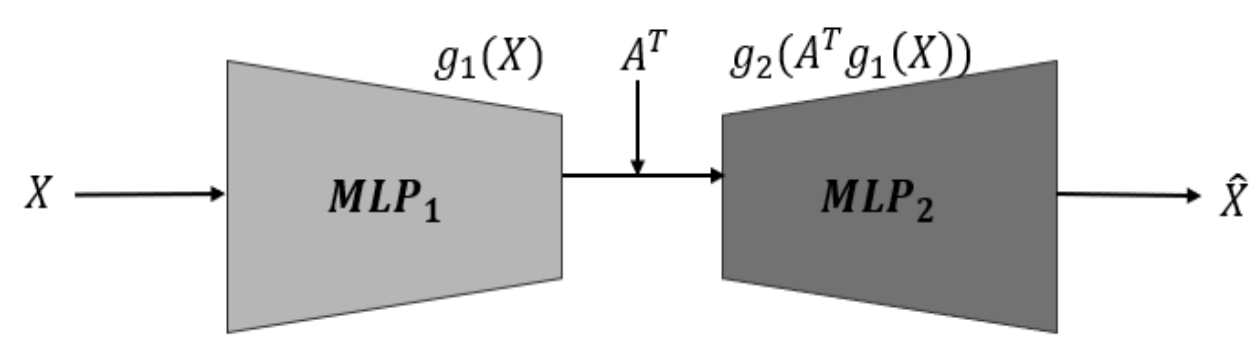

**Fig 3. Graph AE framework. *A*: adjacency matrix**

Thus, the loss function of Graph Autoencoder is computed as follows: (11), (12). Here, the augmented lagrangian method is also used to convert equality constraints induced by conditions of DAG. Unlike other parameters within the Graph Autoencoder structure, which only requires gradient descent algorithms, lagrangian multipliers: $\alpha$ and $\rho$, are both updated based on separate rules as in equation (12).

$$
\begin{aligned}
\min_{A,\theta_1,\theta_2} \mathcal{L}_{GAE} &= \min_{A,\theta_1,\theta_2} \frac{1}{n}\sum_{i=1}^{n} \left\|X^{(i)} - \hat{X}^{(i)}\right\|^2 + \lambda\|A\|_1 + \alpha h(A) \\
&\quad + \tfrac{\rho}{2}|h(A)|^2,\ \text{where } h(A) = tr(e^{A \odot A}) - d \qquad \textbf{(11)}
\end{aligned}
$$

$$
\begin{gathered}
\alpha^{(k+1)} = \alpha^{(k)} + \rho^{(k)} h\left(A^{(k+1)}\right) \\
\rho^{(k+1)} = \begin{cases} \beta\rho^{(k)}, (if\ \left|h\left(A^{(k+1)}\right)\right| \geq \gamma\left|h\left(A^{(k)}\right)\right|) \\ \rho^{(k)}, (o.w.), (\beta > 1, \gamma < 1) \end{cases} \qquad \textbf{(12)}
\end{gathered}
$$

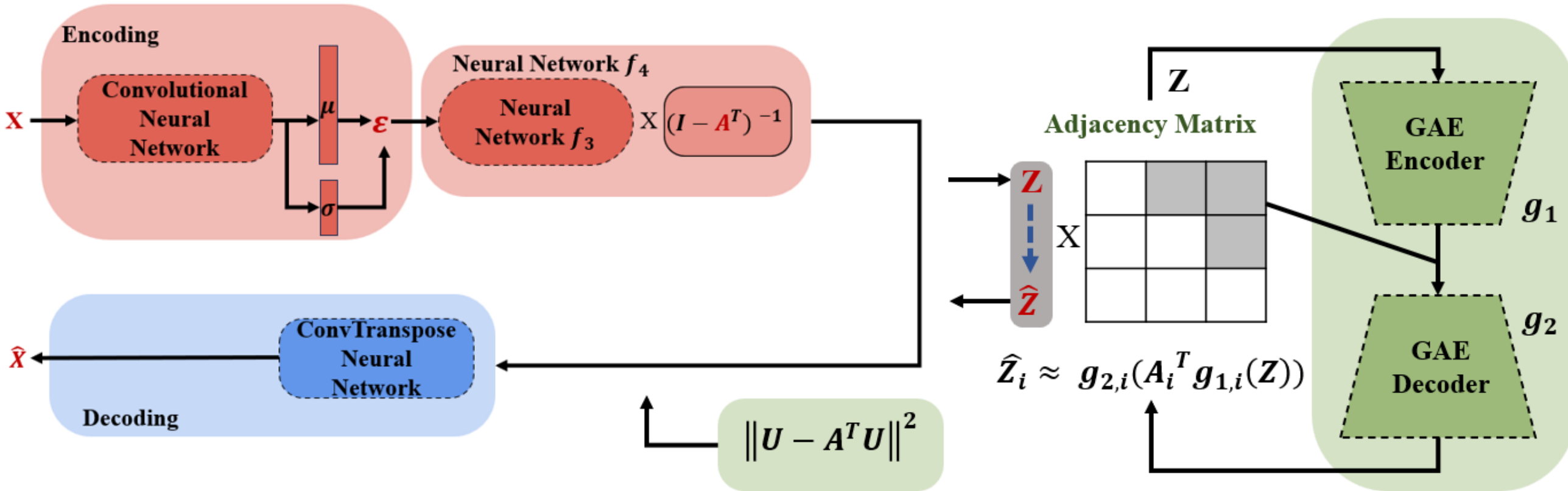

**Fig 4. GCVAMD framework: CausalVAE structure is modified to have higher flexibility and validity by Graph Autoencoder**

*(3) The GCVAMD framework*

GCVAMD, which was designed solely for causal AMD representation learning, combines the structure of CausalVAE and Graph Autoencoder to create a realistic causal flow in multiple aspects. First, as the linearity assumption of SCM used in CausalVAE to compute Z from noise variables $\varepsilon$ (Eq. 4) has high discrepancies with non-linear real world causality, an additional neural network encoder-decoder structure was implemented to compute Z (Eq. 13), which was based on the approach introduced in DAG-GNN [20]. Second, regarding the structural causal masking layer in CausalVAE, though non-linear generative additive modelling(GAM) was implemented in the aspect of $A_i \circ z$, for a more intensive approach, GAM was expanded to parent variables Z itself under the Graph Autoencoder framework introduced in III. Methodology (2) (Eq. 14). Thus, GCVAMD can be considered as a modified version of CausalVAE, which incorporates an encoder-decoder structure in both latent factor extraction from $\varepsilon$ and the structural causal layer. For label information variable $U$, discrete variables: Neovascularization Status, Drusen Status, and the Severity of AMD from OCTDL dataset metadata were implemented in GCVAMD, leading to a causal latent space dimension of $\mathbb{R}^{3\times N}$. A brief graphical abstract of the overall process in GCVAMD is summarized in Fig 4.

$$\boldsymbol{Z} = \boldsymbol{f_4}((\boldsymbol{I} - \boldsymbol{A}^T)^{-1}\boldsymbol{f_3}(\boldsymbol{\varepsilon}))\ \ (non - linear\text{:}\ \boldsymbol{f_{3(4)}}) \qquad \textbf{(13)}$$

$$for\ \forall i^{th}\ variable\ in\ \boldsymbol{Z},\quad \hat{\boldsymbol{Z}}_{\boldsymbol{i}} = g_{2,i}(\boldsymbol{A}^T g_{1,i}(\boldsymbol{Z})) + \boldsymbol{\epsilon_i} \qquad \textbf{(14)}$$

In terms of loss function, to leverage the effect of labels U and to consider different primal objectives of sub-structures in GCVAMD, different $(\beta, \gamma, \nu)$ values and a novel parameter $\omega$ were implemented based on the number of epochs and objects of training(sub-structure) in GCVAMD (Eq 15).

$$GCVAMD\ Loss$$

$$L = \omega\{BCE(p_{ij}, x_{ij}) + \frac{1}{2}\sum {\varepsilon_{ij}}^2\} + \alpha H(A) + \beta l_z + \gamma l_u + \nu \times l_{zu}$$

$$* \ l_u = \sum \|u - A^T u\|^2, l_z = \sum \|z - g_2(A^T g_1(z))\|^2$$

$$H(A) = tr(e^{A \odot A}) - d, A \in \mathbb{R}^{3\times 3} \qquad \textbf{(15)}$$

Moreover, to prevent the occurrence of gradient explosions or vanishing issues, use of separate learning rates based on different types of sub-structures was implemented within training. Specifically, larger learning rates were allocated to weighted adjacency matrix training and GAE weights to ensure causal discovery is emphasized as much as image reconstruction objectives in VAE.

*B. Experimental Design*

As explained in II. Preliminaries, for GCVAMD causality training, 150 AMD retinal OCT images, 150 Normal retinal OCT images and corresponding label data were set as input data. For GCVAMD reparameterization settings, gaussian standard normal distribution was used to generate randomness. For parameter settings in the convolutional encoder and decoder of GCVAMD, values in TABLE I were implemented. Meanwhile, parameters for $\boldsymbol{f_3}$ and $\boldsymbol{f_4}$ (functions that convert noise variables to latent factors Z) were set as TABLE II, whereas parameters in Graph Autoencoder-based Structural Causal Masking layers were set as TABLE III. All sub-structures and weights in GCVAMD were dealt using Tensorflow 2.19.0. For adjacency matrix A, a separate custom weight matrix using .add_weights() function in tensorflow.keras package was implemented for computation. (Initialization of adjacency matrix $A$ was set as zero weights). For label variables $U$, neovascularization status, drusen status, and AMD severity, were each defined as $u_0, u_1, u_2$ sequentially for analysis.

TABLE I
HYPER PARAMETERS SETTINGS 1

| Type | Parameter settings |
|---|---|
| Encoder | 16 Conv2D 5x5 filters, stride=3, silu<br>16 Conv2D 4x4 filters, stride=2, silu<br>32 Conv2D 4x4 filters, stride=2, silu<br>256, FC layer(Flattened), ELU<br>64, FC layer, ELU<br>6(=3+3), FC layer, Linear |
| Decoder | 64, FC layer, ELU<br>256, FC layer(Flattened), ELU<br>17x17x32 FC layer, ELU<br>Reshape(17,17,32) layer<br>16 Conv2DTranspose 4x4 filters, stride=2,silu |
| | 16 Conv2DTranspose 4x4 filters, stride=2,silu<br>3 Conv2DTranspose 5x5 filters, stride=3,sigmoid |

*Silu(Swish): Activation function proposed by [25].

TABLE II
HYPER PARAMETERS SETTINGS 2

| Type | Parameter settings |
|---|---|
| $\boldsymbol{f_3}$ | 3*4(# of nodes), Sparse Layer, ELU<br>3*4, Sparse Layer, ELU<br>3, Sparse Layer, Linear |
| $\boldsymbol{f_4}$ | 3*4, Sparse Layer, ELU<br>3*4, Sparse Layer, ELU<br>3, Sparse Layer, Linear |

*Sparse Layer: Masked FC Layer based on additive modeling approach

TABLE III
HYPER PARAMETERS SETTINGS 3

| Type | Parameter settings |
|---|---|
| GAE encoder | 3*4(# of nodes), Sparse Layer, ELU<br>3*4, Sparse Layer, ELU<br>3, Sparse Layer, Linear |
| GAE decoder | 3*4, Sparse Layer, ELU<br>3*4, Sparse Layer, ELU<br>3, Sparse Layer, Linear |

*Sparse Layer: Masked FC Layer based on additive modeling approach

Masking fully connected layers within GAE: $\boldsymbol{f_3}$, $\boldsymbol{f_4}$ under additive modelling was computed by converting binary mask matrices into kernel constraints with tensorflow's custom class: keras.constraints.Constraint(). When training GCVAMD, a total of 250 epochs with full-batch gradient descent were implemented with first 150 epochs trained based on Algorithm 1, and the last 100 epochs trained under Algorithm 2 (*Initial $\alpha, \rho, \gamma, \beta$ for DAG constraint-based updates explained in (Eq. 12) were set as 0.6, 0.1, 0.9, and 1.01).

**Algorithm 1** Optimization Process in GCVAMD (150 epochs)

**-Initial settings: i = 0 , maximum epoch(K) = 150;**
**-While i < K**:
1. Compute $\{BCE(p_{ij}, x_{ij}) + \frac{1}{2}\sum \varepsilon_{ij}^2\}$, $l_u$, $l_z$, $l_{zu}$ under OCT images
2. Compute $\mathcal{L}_1$= GCVAMD Loss($\omega = 1, \beta = 0.3, \gamma = 0.3, v = 0.1$)
3. Compute $\mathcal{L}_2$ = GCVAMD Loss($\omega = 0.3, \beta = 2.0, \gamma = 0.5, v = 0.1$)
4. Update weights via gradient descent methods:
   1) Update adjacency matrix $A$ and GAE weights with $\mathcal{L}_2$
      *(learning rate: 2e-2, 3e-3)
   2) Update remaining structure weights with $\mathcal{L}_1$
      *(learning rate: 2e-3)
   3) Update $\alpha, \rho$ using **(12)**
5. Update i = i+1;

**Return**: weighted adjacency matrix $A$, Latent Space $Z$

**Algorithm 2** Optimization Process in GCVAMD (100 epochs)

**-Initial settings: i = 0 , maximum epoch(K) = 100;**
**-While i < K**:
1. Compute $\{BCE(p_{ij}, x_{ij}) + \frac{1}{2}\sum \varepsilon_{ij}^2\}$, $l_u$, $l_z$, $l_{zu}$ under OCT images
2. Compute $\mathcal{L}_1$= GCVAMD Loss($\omega = 1, \beta = 0.3, \gamma = 0.3, v = 0.1$)
3. Compute $\mathcal{L}_2$ = GCVAMD Loss($\omega = 0.3, \beta = 2.0, \gamma = 0.5, v = 0.3$)
4. Update weights via gradient descent methods:
   1) Update adjacency matrix $A$ and GAE weights with $\mathcal{L}_2$
      *(learning rate: 4e-2, 3e-3)
   2) Update remaining structure weights with $\mathcal{L}_1$
      *(learning rate: 2e-3)
   3) Update $\alpha, \rho$ using **(12)**
5. Update i = i+1;

**Return**: weighted adjacency matrix $A$, Latent Space $Z$

After fitting procedure was complete, GCVAMD's latent causal space retrieval was evaluated in both quantitative and qualitative aspects. For quantitative evaluations, Total GCVAMD loss and GAE loss convergence, weighted adjacency's Structural Hamming Distance(SHD) when compared to the ground truth adjacency matrix explained in Fig 1, and Disentanglement Score [26] for each Z were used to examine success in fitting, causal structure retrieval, and disentanglements in representations. Here, to compute SHD between causal graphs, converting the achieved weighted adjacency matrix to binary components was computed by setting top 20% weights as 1 and the rest as 0 (magnitude of weights used for ordering). Meanwhile, the Disentanglement Score between causal representations Z and label information U was defined based on (Eq. 16), which is the disentanglement definition induced by C, Eastwood and C.K.I, Williams (2018). The $R_{ij}$ within definition denotes the magnitude of coefficients, or weights under LASSO regressions on each label $U$ [26,27]. Specifically, by definition in [26], GCVAMD latent factors: Z were set as code dimensions, and label information variables: U were set as factors. Thus $R_{ij}$ implies the absolute weight that corresponds to code dimension $z_j$ when factor $u_i$ is defined as the response variable in LASSO. For LASSO regression implementations, Lasso() function within sklearn.linear_model package with α set as 0.1 (for $u_0, u_1$ prediction) and 0.01 (for $u_2$ prediction) was used in Disentanglement Score computations. (*To avoid computation errors, if LASSO's coefficient was reduced to 0, a sufficiently small constant of $\epsilon$=1e-5 was added to 0 for computation).

$$\text{Let } prob\ p_{ij} \text{ in } D_j \text{ denote } p_{ij} = \frac{R_{ij}}{\sum_{k=0}^{M-1} R_{kj}}$$
$$\Rightarrow Modularity\ for\ z_j: D_j = 1 - \sum_{i=0}^{M-1} p_{ij} \log_M \frac{1}{p_{ij}} \quad \textbf{(16)}$$

(*All experiments and model constructions were based on Python Tensorflow 2.19.0. All procedures were implemented under Google Colab's basic T4 NVIDIA GPU environments).

For qualitative disentanglement evaluations, visualization based on varying individual latent causal factor: Z while others being fixed was implemented to check disentanglements. Based on candidate sets: {-0.2, -0.1, -0.05, 0, 0.05, 0.1, 0.2}, {-0.1, -0.05, -0.025, 0, 0.025, 0.05, 0.15}, and {-0.1, -0.05, 0, 0.05, 0.1, 0.2, 0.3}, changes in reconstructed images based on varying $z_0$ to $z_3$ were sequentially checked to evaluate correspondence with neovascularization, drusen and AMD status.

### C. AMD Prediction using Latent Causal Factors

After successful causal retrieval and disentanglement were achieved in GCVAMD, this research extended latent causal factors to downstream tasks. Specifically, causal GCVAMD encodings: $z_0, z_1$ that correspond with major causes: neovascularization and drusen, and $\hat{z}_2$, which is computed from $z_0, z_1$ and adjacency matrix $A$, were used in detecting AMD from raw OCT images. To check whether latent causal information from GCVAMD improves conventional non-causal neural network approaches, a baseline convolutional Autoencoder, which was used to extract non-causal latent features, and a simple 4-layer DNN, which was used to classify Normal/AMD retinas, were implemented for comparison. By comparing AMD detection results from convolutional autoencoder-based latent features and causality incorporated features ($z_0, z_1, \hat{z}_2$+ convolutional AE-based latent features), significance of GCVAMD usage in downstream prediction tasks was further validated. For training data, identical training set, which was implemented for GCVAMD causal fitting, was used for fitting. Meanwhile, for test data, 182 normal retina OCT images, which were left after GCVAMD training set definition, and 182 random sampled AMD diagnosed retinal OCT images from non-training datasets (AMD image) were used to evaluate performance in AMD detection (*number of AMD test images matched with normal retinal images)

Parameters for convolutional AE and DNNs were set as TABLE IV and TABLE V. Learning rates and batch size for fitting under Adam optimizers were set as (5e-3, 100) (AE), and (1e-4, 50) (DNN). For convolutional AE fittings, a total of 300 epochs were used for fitting, whereas 400 epochs were used for DNNs. For classification metrics, five metrics: classification accuracy, precision score, recall score, Macro f1-score, and ROC-AUC scores were implemented based on test data.

TABLE IV
HYPER PARAMETERS SETTINGS IN CONVOLUTIONAL AE

| Type | Parameter settings |
| --- | --- |
| Encoder | 16 Conv2D 5x5 filters, stride=3, silu<br>16 Conv2D 4x4 filters, stride=2, silu<br>32 Conv2D 4x4 filters, stride=2, silu<br>256 FC layer(Flattened), ELU<br>64 FC layer, ELU<br>10 FC layer, Linear (Latent layer) |
| Decoder | 10 FC layer, ELU<br>64 FC layer, ELU<br>256 FC layer(Flattened), ELU<br>17x17x32 FC layer, ELU<br>Reshape(17,17,32) layer<br>16 Conv2DTranspose 4x4 filters, stride=2,silu<br>16 Conv2DTranspose 4x4 filters, stride=2,silu<br>3 Conv2DTranspose 5x5 filters, stride=3,sigmoid |

TABLE V
HYPER PARAMETERS SETTINGS IN DNN

| Type | Parameter settings |
| --- | --- |
| DNN | Batch Normalization( )<br>32(# of nodes), FC layer, ELU<br>Batch Normalization( )<br>16, FC layer, ELU<br>Batch Normalization( ) |

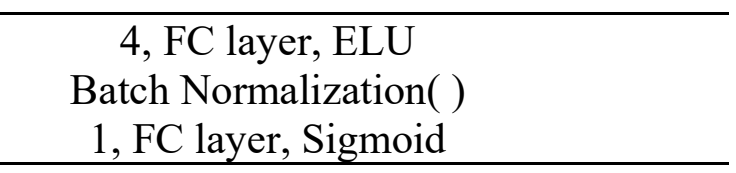

| | |
| --- | --- |
| | 4, FC layer, ELU<br>Batch Normalization( )<br>1, FC layer, Sigmoid |

*Input Dimension: (182, 13) or (182,10)

## IV. EXPERIMENTAL RESULTS

### A. GCVAMD fitting and Causal Retrieval Results

Results of latent causal representation learning with GCVAMD were deduced as follows. For loss values, $\mathcal{L}_1$ =0.6882, $\mathcal{L}_2$ =0.5814, $BCE(p_{ij}, x_{ij}) + \frac{1}{2}\sum \varepsilon_{ij}{}^2 =$ 0.5177, Graph AE loss($l_z$)= 0.0057, label loss ($l_u$) = 0.1723 were deduced under stable convergence (Fig 5). Thus, successful fitting and latent causal space retrieval was checked under OCT images and label data.

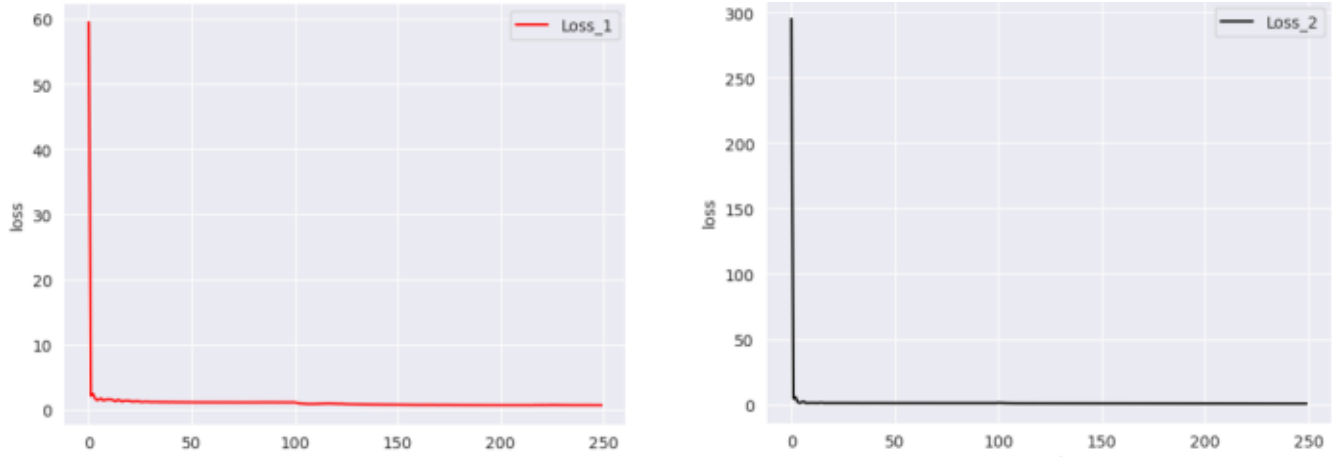

**Fig 5. Visualization of loss values from fitting GCVAMD model. (Left) $\mathcal{L}_1$ per epoch (Right) $\mathcal{L}_2$ per epoch**

Next, extracted latent causal adjacency matrix was converted to binary components as Fig 6. Comparing binarized adjacency matrix by GCVAMD and the ground truth AMD graph in Fig 1., SHD of 1.0 was deduced, which implies that extracted causal graph achieved high significance in terms of the domain causal structure correspondence. Thus, latent causal space from GCVAMD was found to have high possibility of representing AMD causal factors and severity of AMD with validity.

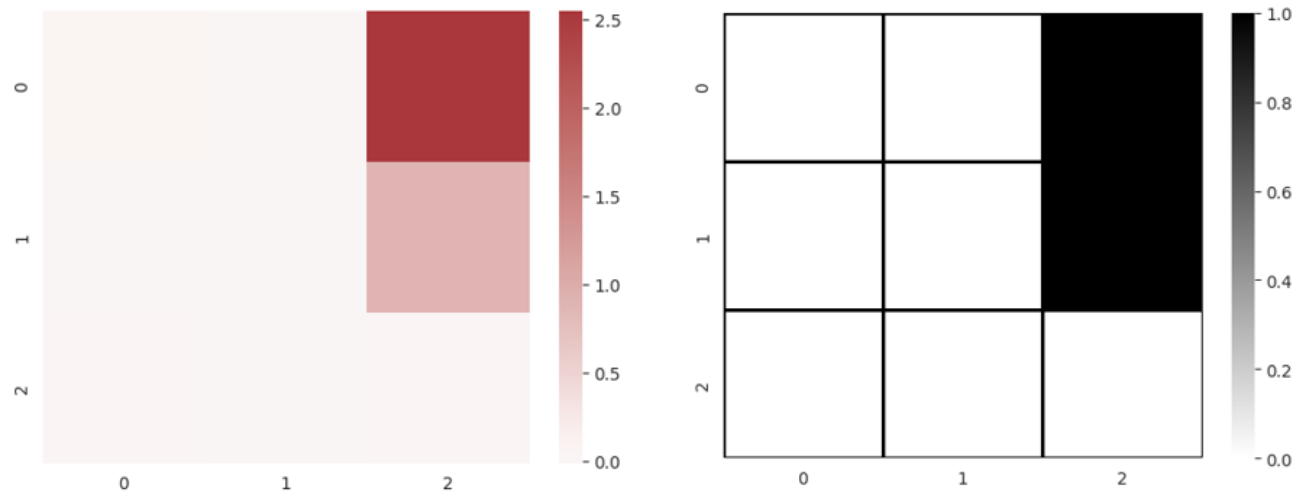

**Fig 6. (Left) Latent adjacency matrix (Right) Binarized *A*.**

### B. Causal disentanglement checks

Under successful latent causal adjacency extraction from GCVAMD, disentanglements regarding variables: $Z_0$ to $Z_2$ were checked to validate whether extracted latent variables can be one-to-one matched with domain causal factors within AMD mechanism. First, for quantitative evaluation, disentanglement score $D_i$ was computed based on LASSO regressions. LASSO regression based $p_{ij}$ computation results were deduced as TABLE VI. Using (Eq 16.) from [26,27], modularity, or disentanglement scores for variables $Z_0$ to $Z_2$ were computed as $D_0$= 0.3697, $D_1$= 0.1619, $D_2$= 0.7573. Having a moderate mean disentanglement score of 0.4296, latent variables Z from GCVAMD were found to be significant in disentangling AMD features: neovascularization, drusen and AMD severity. (*Though $D_1$ was found to be relatively smaller than other latent factor disentanglement scores, considering high label imbalance in drusen status meta data, it was presumed that $D_1$ also indicates partial success in representations within research).

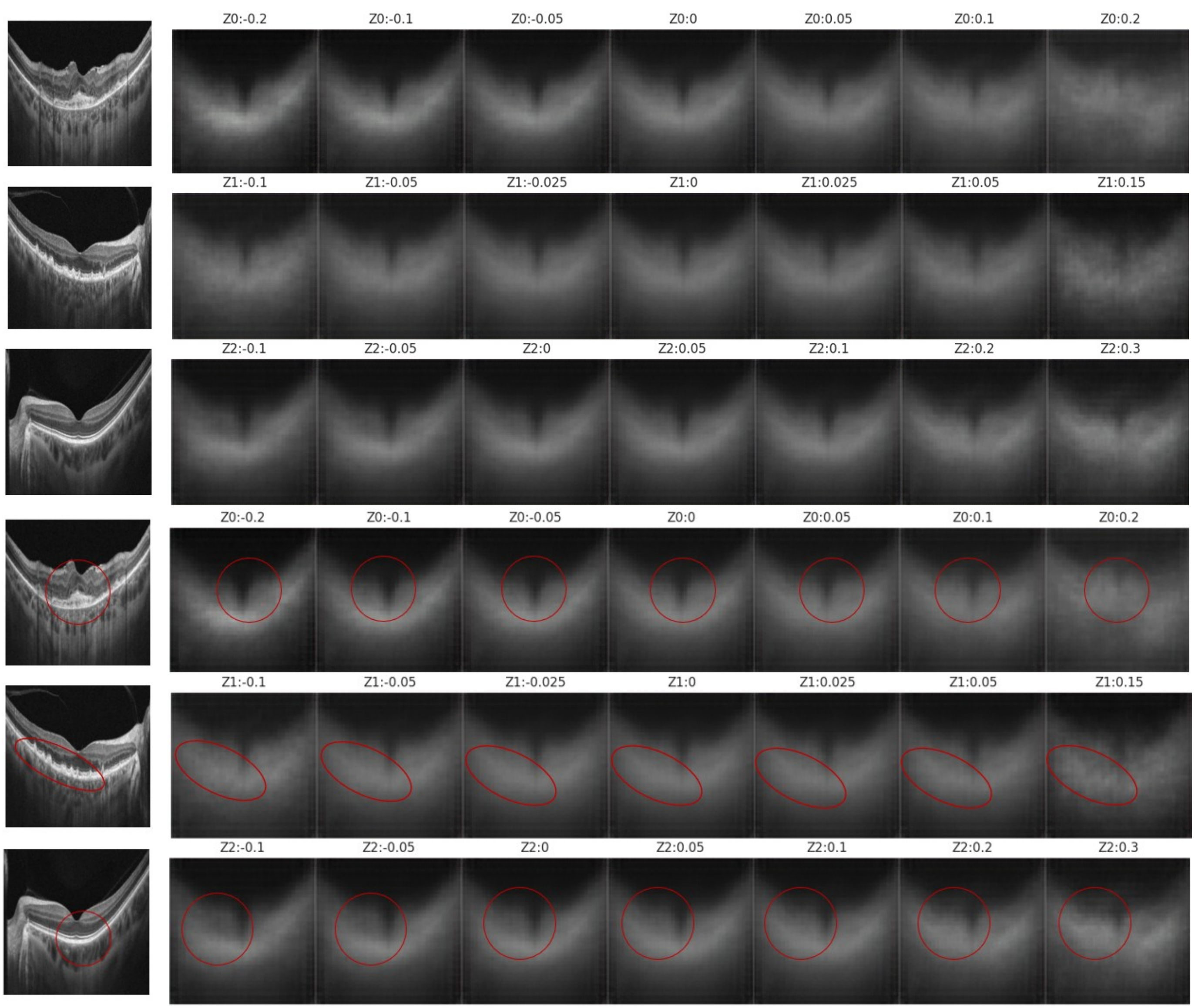

**Fig 7. Examples of visualization of varying latent causal representations(factors) Z0 to Z2 (sequentially). Red circles in 4th to 6th rows are an emphasis on changes within generated OCT retinas (first to third rows) by GCVAMD.**

TABLE VI
$p_{ij}$ MATRIX BASED ON LASSO REGRESSION UNDER $\alpha = 0.1, 0.01$

| $p_{ij}$ values | $v_0(\equiv u_0)$ | $v_1(\equiv u_1)$ | $v_2(\equiv u_2)$ |
|---|---|---|---|
| $Z_0$ | 0.47914 | 0.43130 | 0.07422 |
| $Z_1$ | 0.00002 | 0.08340 | 0.00024 |
| $Z_2$ | 0.52085 | 0.48530 | 0.92554 |

Second, for qualitative AMD risk factor disentanglement checks, random visualization results of varying individual latent causal factors within GCVAMD image reconstructions were implemented with other variables being fixed. Examples of variation in causal latent variables: $Z_0, Z_1, Z_2$, which were each found to correspond with neovascularization status, drusen, and AMD severity, were visualized in Fig 7 and Fig 8, sequentially. (First column denotes original image: AMD with neovascularization, AMD with drusen, Normal Retina (sequentially)). For variable $Z_0$, as $Z_0$ values were increased from -0.2 to 0.2, generation of hyper reflective mass around the RPE layer, and an overall uplift in the retina was detected, which is a representative feature of choroidal neovascularization (CNV) within the retina [28]. Considering domain features of CNV(hyper-reflectivity and protrusions

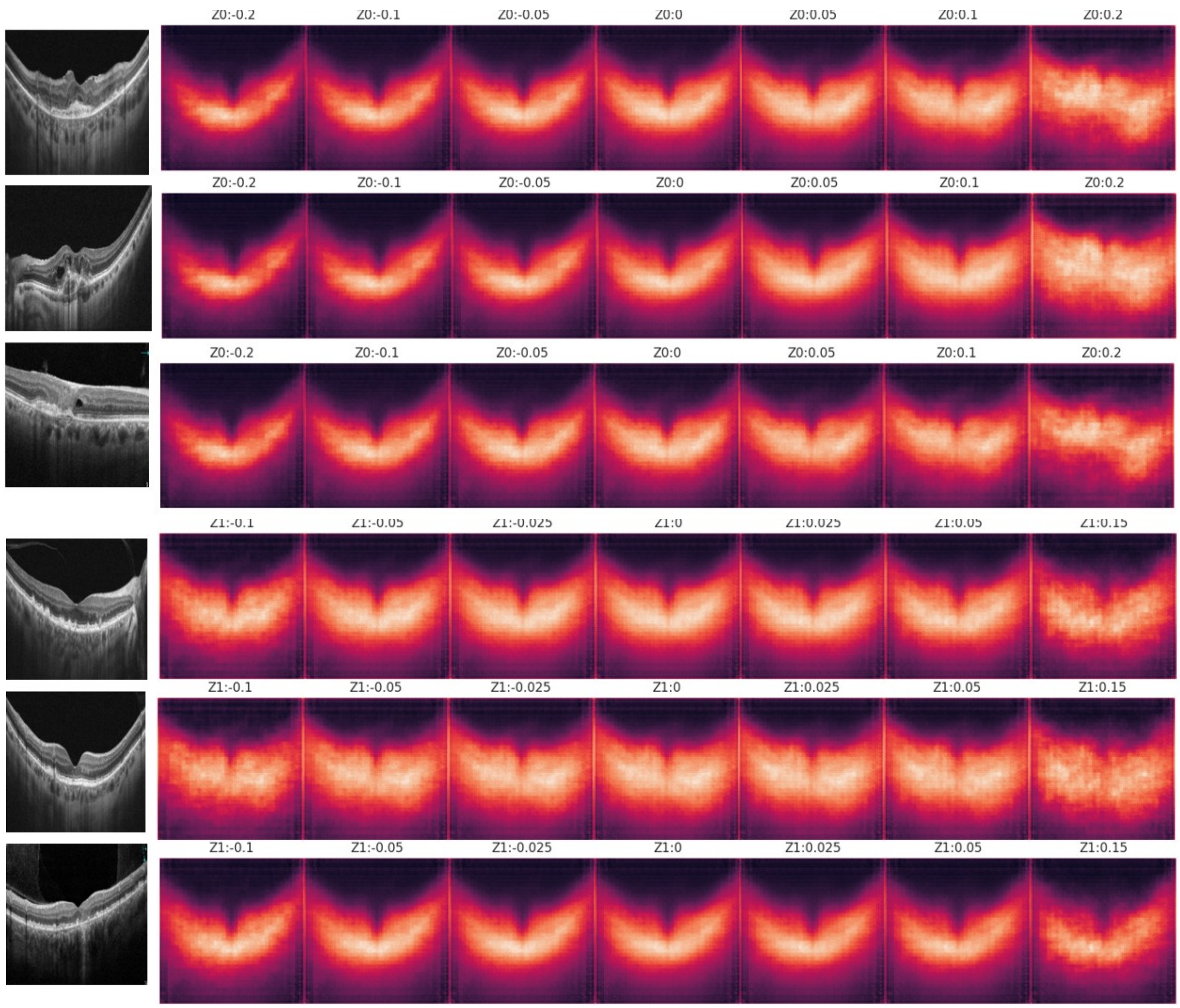

**Fig 8. (Continued from Fig 7.) Examples of VAE-based reconstructed images (single channel) from varying latent causal representations(factors): Z0 and Z1. First to Third row: Varying Z0, Original Image sampled from neovascularization-based AMD cases. Fourth to Sixth row: Varying Z1, Original Image sampled from drusen-based AMD cases. More vivid changes within reconstructed images are visible when varying individual latent variables: Z0 and Z1**

towards the vitreous [29]) and the region of neovascularization in the original image, it was found plausible to conclude that $Z_0$ can one-to-one correspond with CNV with high significance.

Meanwhile, for variables $Z_1$, as values deviated from 0, undulation in the original RPE region was highly increased. That is, the smoothness of RPE layers within the reconstructed image was deteriorated as $Z_1$ deviated from 0. Considering domain features of drusen(RPE elevation due to mounds of deposits, saw tooth shaped patterns near the RPE layer [30,31]) and the region of drusenoid deposits in the original image, it was found plausible to conclude that $Z_1$ has correspondence with drusen status with high significance. This is a significant finding, as extreme imbalance in drusen status labels(non-drusen: drusen = 9.7 : 1) was present within training data.

Lastly, for variables $Z_2$, as values increased from -0.1 to 0.3, creations of hyperreflective mass and noise along the RPE layer were found. Furthermore, uplifting in the RPE layer was also checked when $Z_2$ was increased from 0 to 0.3. Based on the fact that the original image is a representative normal OCT image of the retina, and the fact that subretinal hyperreflective material(SHM) or hyperreflective foci are direct features of AMD, it was found plausible to conclude that $Z_2$ can one-to-one correspond with severity of AMD among patients with high significance.

Thus, through fitting results of GCVAMD and disentanglement checks in latent causal factors, it was found that the GCVAMD can identify and extract AMD causal mechanisms with validity. This implies that, based on the concept of Structural Causal Layers in CausalVAE [11], it is possible to compute valid intervention analysis or simulations of treatments on major AMD factors by intervening individual factors: $z_i$ in Graph Autoencoder inputs within GCVAMD. Not only enabling AI models to comprehend causal structures within AMD, but also enabling intervention analysis, GCVAMD can efficiently help current AMD detecting AI-models to overcome analytic limitations induced by correlation-dependent approaches.

### C. AMD Detection using Latent Causal Factors

Under success in AMD latent causal space retrieval and disentanglement validation, enhancements in downstream task performance when incorporating GCVAMD encodings: $Z_0$, $Z_1$, $\hat{Z}_2$ were checked under experimental settings in III. AMD classification results from test data were summarized in TABLE VII and (Fig 9, Fig 10) (Causality Incorporated: $Z_0$, $Z_1$, $\hat{Z}_2$ concatenated to latent vector of AE, Not Incorporated: only latent vector of AE is used). In all evaluation metrics except for precision score, causality incorporated DNN showed significantly higher performance within test data. Specifically, causality incorporated DNN showed 75.27% accuracy, 75.25% f1-score, and 83.25% ROC AUC score, whereas the non-incorporated DNN exhibited 72.53% accuracy, 72.26% f1-score, and 77.20% ROC AUC score. Thus, causality incorporated DNN achieved 2.74%p higher accuracy, 2.99%p higher f1-score, and 6.05%p higher ROC AUC scores in AMD classification tasks. Furthermore, causality incorporated DNN also exhibited a 9.34%p higher recall score. This implies that though concatenating only three AMD causal representations from GCVAMD to conventional CNN-based approaches, significant increase in AMD classification performance can be achieved in downstream tasks.

TABLE VII
AMD CLASSIFICATION PERFORMANCE ON TEST DATA

| Type | Causality Incorporated | Not Incorporated |
|---|---|---|
| Accuracy | **75.27%** | 72.53% |
| Precision | 77.06% | **78.08%** |
| Recall | **71.98%** | 62.64% |
| F1-score | **75.25%** | 72.26% |
| ROC AUC | **83.25%** | 77.20% |

**higher metric value emphasized in bold*

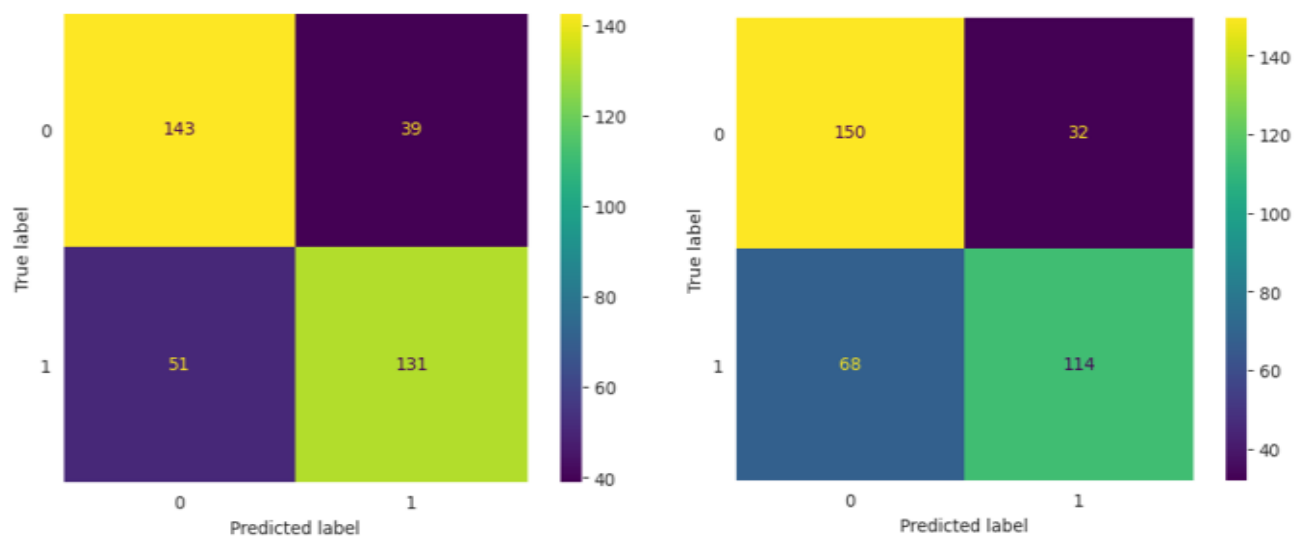

**Fig 9. Confusion Matrix from AMD classifications. (Left) Causality incorporated (Right) Not incorporated**

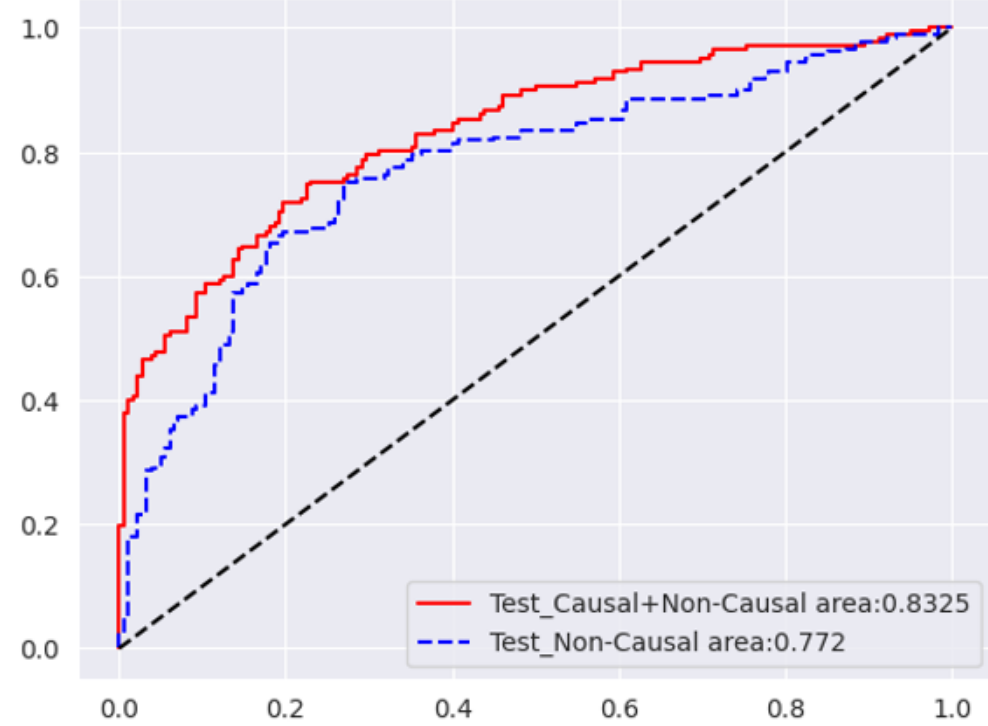

**Fig 10. Visualization of ROC AUC curves from each DNN (Causality informed, Not informed).**

## V. CONCLUSION

In this research, a novel causal AI framework for Age-related Macular Degeneration (AMD) analysis: GCVAMD was constructed and was validated in terms of qualitative, quantitative evaluation metrics, and AMD detection downstream task performance enhancements. Specifically, by modifying the CausalVAE framework from [11] with Graph Autoencoder approaches in causal layer masking, non-linearity assumptions for computing latent variables from noise variables, and alleviation of supervision regarding label data, successful identification of major AMD causes: neovascularization and drusen, and even severity of AMD status was achieved through two-step training under OCTDL datasets. Furthermore, when varying individual AMD latent causal factors (with other variables being fixed), reconstructed images or generated images from GCVAMD exhibited representative features of corresponding AMD-related features, which implies that causal intervention analysis on specific factors can be driven not only in terms of quantitative aspects, but also in visual aspects. As previous Computer Vision based AMD researches were mainly focused on classification tasks or segmentation itself with only correlation information on hand, it is strongly believed that the proposed framework: GCVAMD can significantly enhance AMD analysis by incorporating causality into current AMD treatment approaches or existing models. Furthermore, as causal identification of major features within the AMD mechanism was checked, more reliable detections or classification of AMD in early stages can be expected by incorporating latent vectors: $Z_0$, $Z_1$, $\hat{Z}_2$ to downstream tasks.

However, there are also several limitations which should be dealt for more extensive use of GCVAMD in various fields. First, due to high label imbalance in drusen based AMD data (No: Yes = 9.7 : 1), identification of drusen in latent space is relatively weak, when compared to other components within AMD's causal graph. Second, the ground truth AMD causal graph within this work may have missed significant mediators, such as hemorrhage or sub-retinal fluids in the retina. Third, as the OCTDL dataset does not consider cases where drusen and neovascularization are both present within the patient's eye, some labels within data could be insufficient in representing AMD status of OCT images, which can hinder optimal GCVAMD fitting. (Though using an alleviation approach under a weighted MSE was implemented in this work to deal with label insufficiency, it would be more ideal to acquire full meta data that contains sufficient information regarding AMD factor-wise conditions).

Presuming that these limitations being dealt, GCVAMD is believed to have enough significance to be applied to early AMD treatments in practical aspects, which can extremely reduce the possibility of acute vision impairment due to AMD.